\documentclass{aa}
\usepackage{epsfig}
\begin{document}
\newcommand{\be}{\begin{eqnarray}}
\newcommand{\ee}{\end{eqnarray}}
   \thesaurus{06(08.14.1; 08.05.3; 08.13.1)}
   \title{Neutrino-pair emission in
             a strong magnetic field}
   \subtitle{}
   \author{E.N.E. van Dalen, A.E.L. Dieperink,
           A. Sedrakian, R.G.E. Timmermans}

   \institute{Kernfysisch Versneller Instituut,
         NL-9747 AA Groningen,
         The Netherlands\\
              email: userlastname@kvi.nl}
 \date{received: \ 16 February 2000 \ ;accepted:}
  \authorrunning{van Dalen et al}
   \titlerunning{neutrino-pair emission}
   \maketitle
   \begin{abstract}
   We study the neutrino emissivity of strongly magnetized
   neutron stars due to the charged and neutral current couplings
   of neutrinos to baryons in strong magnetic fields. The leading
   order neutral current process is the one-body neutrino-pair
   bremsstrahlung, which does not have an analogue in the zero field
   limit. The leading order charged current reaction is the known
   generalization of the direct Urca processes to strong magnetic
   fields. While for superstrong magnetic fields in excess of $10^{18}$ G
   the direct Urca process dominates the one-body bremsstrahlung, we
   find that for fields on the order $10^{16}-10^{17}$ G and temperatures
   a few times $10^9$ K the one-body bremsstrahlung is the dominant process.
   Numerical computation of the resulting emissivity, based on a simple
   parametrization of the equation of state of the $npe$-matter in a
   strong magnetic field, shows that the emissivity of this reaction
   is of the same order of magnitude as that of the modified Urca process
   in the zero field limit.

   \end{abstract}
\section{Introduction}
It is now well established that the neutron stars which
are observed as radio-pulsars posses $B$-fields of
the order of $10^{12}$-$10^{13}$ G at the surface.
The interior fields are unknown, but can be
by several orders of magnitude larger than the ones
inferred for the surface.
The scalar virial theorem
sets an upper limit on the magnetic field strength of
a neutron star of the order of $10^{18} \rm{G}$
(Lai \& Shapiro \cite{Lai}).
Similar conclusion is reached through more
sophisticated  numerical studies (Bocquet et al. \cite{Boc}).

Recent measurements of the spin-down  timescales
of several soft gamma-ray repeaters,
such as  SGR 0526-66 (Mazets et al. \cite{Maz}), SGR 1806-20
(Murakami et al. \cite{Mur}), and SGR
1900+14 (Kouveliotou et al. \cite{Kou1}) with
RXTE, ASCA and BeppoSAX have made a strong case for
SGRs as being newly born neutron stars that have
very large surface magnetic fields (up to $10^{15}$ G).
 The subsequent discoveries of the SGR 1627-41
by BATSE (Woods et al. \cite{Woods})
and  SGR 1801-23  by Ulysses, BATSE, and KONUS-Wind (Cline et al. \cite{Cline})
lent further support to the identification
of SGRs with highly magnetized neutron stars.
These objects were naturally related to the magnetars, which are
thought to be remnants of a supernova explosion which develop high
magnetic fields via a dynamo mechanism
(Duncan \& Thompson \cite{Dun}, Thompson \& Duncan \cite{Thom}). The
magnetars also
serve as a model for the
anomalous X-ray pulsars (AXP) (van Paradijs, Taam \& van den Heuvel \cite{Para})
such as 1E 1841-045 (Kes 73) (Gotthelf, Vasisht, \& Dotani \cite{Gott}),
RX J0720.4-3125 (Haberl et al. \cite{Hab}), and 1E 2259+586
(Rho \& Petre \cite{Rho}).

Neutrino-nucleon interactions in the strong magnetic fields
have been studied recently both in the supernova and neutron
star contexts. It has been pointed out that the
neutrino emission from proto-neutron stars, which
is anisotropic in strong $B$-fields, could produce the
``pulsar kicks"  if the fields  are in excess of
$10^{16}$ G (Horowitz \& Li \cite{Horo};
Arras \&  Lai \cite{Arras} and references therein).
The strong magnetic fields relax the kinematical constrains
on the direct Urca process and hence give rise to finite
neutrino emissivity even when the proton fraction is small
(Leinson \&
Perez \cite{Leinson},  Bandyopadhyay et al. \cite{Band}).
The effect is most pronounced in the ultra-high magnetic fields
when  protons and the electrons occupy the lowest Landau levels.
The direct Urca process for arbitrary magnetic fields, when the protons and
electrons are allowed to occupy many Landau levels, has also
been studied (Baiko \& Yakovlev \cite{Baiko}).

The neutrino emissivities via the one-body processes sensitively
depend on the abundances of the various species of baryons and leptons
which are controlled by the equation of state (EoS) of the dense matter
in strong magnetic fields. The strong magnetic fields lead to an increase
of the proton fraction (Broderick, Prakash \& Lattimer \cite{Bro}).
The  muon production and pion condensation in strong magnetic
softens the EoS of the dense matter (Suh \& Mathews \cite{Suh}).
For the purpose of estimating the magnitude of the neutrino emissivities,
we employ in this paper a simple parametrization of the EoS for the
$npe$-matter in strong magnetic fields.

The main objective of this paper is to show that the
strong magnetic fields open a new channel of neutrino emission
via one-body neutral current bremsstrahlung, which
does not have an analogue in the zero
field limit.
We also briefly discuss the direct Urca process, which is forbidden
in the low-density zero-field limit (as long as the
triangular condition $p_{Fp}+p_{Fe}\ge p_{Fn}$ is not satisfied), but is allowed in
strong magnetic fields because of the relaxation of the kinematical
constrains.
We compare the emissivities
of various reactions using a simple parametrization of the
EoS for the $npe$ matter in a strong $B$-field.
As a standard reference for our comparison we
use the modified Urca process.
The presence of a magnetic field has two
different effects on neutrino emissivities:
(i) in pure neutron matter it allows for spin-flip
transitions
where the finite  difference of the momenta of neutrons at two
different Fermi surfaces enables one to
satisfy energy-momentum conservation,
(ii) the charged particles occupy Landau levels
leading to a smearing of the transverse
momenta over an amount $\sqrt{eB}$
\footnote{We use the natural units,  $\hbar=c=k_B=1$.}.
As a consequence there are two typical scales
of the magnetic field,
where effects on the emissivity are expected; first for
  $|\mu_B| B \sim T$ which is relevant in neutrino-pair bremsstrahlung
  from neutrons,
  and second $e B \sim p_F^2$  relevant for the Urca process.

As well know, the emissivity of  any particular reaction
can be related  to the imaginary part of the
polarization function of the medium (Voskresensky \&
Senatorov \cite{VS}, Raffelt \& Seckel \cite{Raffelt}, Sedrakian \& Dieperink \cite{Se1}).
We compute the polarization function of neutrons
and protons in strong magnetic fields employing the
finite temperature Matsubara Green's functions technique.
For the case of the bremsstrahlung  the time-like properties of the
polarization function are relevant. The space-like properties of the
polarization function, relevant for the neutrino-nucleon scattering,
have been studied by Arras \& Lai (\cite{Arras}) in an equivalent
response function formalism.

The plan of this paper is as follows.
The bremsstrahlung emissivity is related to the
polarization function of the medium in Sect. 2.
The neutrino emissivity via neutrino-pair bremsstrahlung
from neutrons is discussed in Sect. 3 and
that from protons in Sect. 4.
The Urca process in strong magnetic fields is briefly discussed Sect. 5.
The EoS of $npe$ matter in a  magnetic field is discussed in Sect. 6.
Our numerical results are presented in Sect. 7.
Sect. 8 contains our conclusions.
\section{The bremsstrahlung emissivity}
The neutrino emissivity of an infinite medium of
interacting hadrons can be expressed in terms of
the imaginary part of the finite temperature polarization
function $\Pi(\vec{q},\omega),$
where $\vec{q}$ and $\omega$ denote the momentum and energy
transfer to the leptons.
In the single-loop approximation the finite temperature polarization function
in a magnetic field is a 2 $\times $ 2 matrix in spin space (Mattuck \cite{Mat})
\begin{eqnarray}
  \Pi_{s,s'} (\vec{q},\omega)= \frac{-2}{\beta} \int \frac{d^3p}{(2
\pi)^3}
\sum_{i(p+q)}
G_{s'}( \vec{p} +
\vec{q} , ip+iq) \nonumber\\
G_s( \vec{p} ,ip),
\label{generalpolarization}
\end{eqnarray}
where $\beta$ is the inverse
temperature. The single-particle Green function can be expressed as
$G_s(\vec{p},ip)=[ip-(E_{p,s}- E_F)]^{-1}$, where $E_F$ is
the Fermi energy and  $ip$ is the Matsubara frequency;
 $s',s=\pm 1 $ specify the initial and final nucleon spins.
(We assume the magnetic field along the positive $z-$axis.)
Carrying out the frequency summation
and taking the imaginary part one finds the well-known result
(Mattuck \cite{Mat})
\begin{eqnarray}
{\rm Im} \Pi_{s,s'}(\vec{q},\omega) = 2 \pi \int \frac{d^3p}{(2
\pi)^3}
(f(E_{p,s})-f(E_{p+q,s'})) \nonumber \\
\delta(E_{p,s}-E_{p+q,s'}+\omega),
\label{polarization}
\end{eqnarray}
where $f(E_{p,s})=[e^{\beta
(E{p,s}-E_{F})}+1]^{-1}$ is the Fermi-distribution function.
The emissivity is then given by
\begin{eqnarray}
\epsilon_{\nu \overline{\nu}}
=3 \sum_{s s'} |M_{s,s'}|^2 \int \frac{d^3p_{\nu}}{(2 \pi)^3}
\frac{d^3p_{\overline{\nu}}}{(2\pi)^3}
d\omega \, d^3q \, \, {\rm Im} \Pi_{s,s'}(\vec{q},\omega)  \nonumber\\
 g_B(\omega) \, \omega \,
\delta({\omega}-{\omega}_{\nu}-{\omega}_{\overline{\nu}}) \,
\delta^3(\vec{q}-\vec{p}_{\nu}-\vec{p}_{\overline{\nu}}) ,
\label{emiss}
\end{eqnarray}
where $g_B(\omega)=[e^{\beta \omega}-1]^{-1}$ is the Bose distribution
function, the factor 3
comes from the sum over  the neutrino flavours, and
the weak interaction matrix (neglecting lepton momenta)
 is
 $|M_{s,s'}|^2=(G_F/2)^2 (c^2_V \delta_{s,s'} + c^2_A
 (2 \delta_{s,-s'} + \delta_{s,s'}))$ with
$c_V$ and $c_A$ the vector and axial vector coupling constants
and $G_F$ the Fermi weak
coupling constant (see Appendix B).
\section{Neutrino-pair bremsstrahlung from neutrons \
($n \rightarrow n + \nu + \overline{\nu}$) }
{ In the absence of a magnetic field
the imaginary part of the polarization function  vanishes for time-like processes
in the quasi-particle approximation,
because energy and momentum cannot
be conserved simultaneously.
In a pure neutron system  a  magnetic field $B$ will give rise to a difference
between the Fermi momenta
of the neutrons with spin
parallel and spin anti-parallel to the $B$-field
(see Appendix A)
\begin{displaymath}
 p^{s}_{Fn}=(E^2_{Fn}-m^2 -s g_n e B)^{1/2},
\end{displaymath}
 where $\mu_n=g_ne/(2m)$ is the neutron magnetic moment
 and $g_n=-1.91$ is the neutron $g$-factor and we assume
 $\mu_n B \ll m$. For $B \not= 0$ energy-momentum conservation
 can be satisfied,} and
as a result one may expect that a field with  strength $| \mu_n | B
\sim T$  leads to a finite
spin-flip polarization function whenever $ \omega
\sim |\mu_n| B$.
\subsection{Polarization function}
In the time-like region it is preferable
to use the relativistic kinematics. The non-relativistic
kinematics does not produce
the correct zero-field limit  $\Pi(\vec{q},\omega,B=0)=0$,
rather a spurious finite contribution
$\Pi(\vec{q},\omega,B=0) \propto \exp(-m/T)$.
To evaluate the angular integral in eq. (\ref{polarization})
note that the energy
conserving $\delta$-function is non-zero if
$|\cos(\theta_{pq})| \le 1$, where $\theta_{pq}$ is
the angle between the momentum vector $\vec{p}$ of the
neutron and the momentum transfer vector $\vec{q}$;
this condition yields the minimal and maximal values of the
three-momentum of neutrons ($p_{\rm{min}}$ and $p_{\rm{max}}$)
for which the imaginary part of the polarization function
is finite.

To obtain a real solution for $p_{\rm{min}}$ and $p_{\rm{max}}$
using the relativistic energy/momentum relation,
$E^2_{p,s}=m^2+p^2- s g_n e B$, the conditions
 $a \equiv [(4 m \mu_n B+\omega^2-q^2)^2-4(m^2+2 m \mu_n B)(\omega^2-q^2)]>0$
and $s'=-s=1$
should be fulfilled.
The result is
\begin{eqnarray}
p_{ {\rm min} }(q,\omega) =\frac{|(-4 m \mu_n B-\omega^2+q^2)q-\omega \sqrt{a}|}{2(\omega^2-q^2)}
,
\end{eqnarray}
and
\begin{eqnarray}
p_{{\rm max}}(q,\omega) =
\left\{ \begin{array}{cc}
\frac{[-4 m \mu_n B-\omega^2+q^2]q+\omega \sqrt{a}}{2(\omega^2-q^2)}
 & \ \ \ \ \ \ b \geq 0 \\
\infty & \ \ \ \ \ \ b < 0
\end{array} \right.
\end{eqnarray}
with $b \equiv 2 E_{p,s} \omega + 4 m \mu_n B+\omega^2-q^2$
(note that $\mu_n < 0 $).
%
Replacing  the integral over the absolute value of the momentum $p$ by
an energy integral $dE/dp=p/E
\approx p/m$ yields
\begin{eqnarray}
{\rm Im} \Pi_{s,s'}(\vec{q},\omega)= \frac{m}{q}
 \int_{p_{\rm{min}}}^{p_{\rm{max}}} \frac{p dp}{2 \pi}
(f(E_{p,s})-f(E_{p+q,s'}))  \nonumber\\  \delta_{s',1} \delta_{s,-1}
=  \frac{m^2}{2 \pi q \beta}
\Bigg(\ln \frac{e^{-\beta E_{\rm{min}}}+1}
 {e^{-\beta (E_{\rm{min}}+\omega)}+1} \nonumber\\
 -\ln \frac{e^{-\beta E_{\rm{max}}}+1}
 {e^{-\beta (E_{\rm{max}}+\omega)}+1} \Bigg) \delta_{s',1} \delta_{s,-1}
\label{polarization2}
\end{eqnarray}
with
$E_{\rm{min/max}}=(m^2+p^2_{\rm{min/max}}- s g_n
Be)^{1/2} - E_{Fn} $.
\\ For large $\beta$ the rhs of Eq. (\ref{polarization2}) is non-negligible only if
$E_{\rm{min}}< 0$ and $E_{\rm{max}}> 0$. The latter condition
is of minor importance, since it is satisfied for almost all $q$ and $\omega$. \\
\begin{figure}
\begin{center}
\includegraphics[angle=0,width=0.45\textwidth]{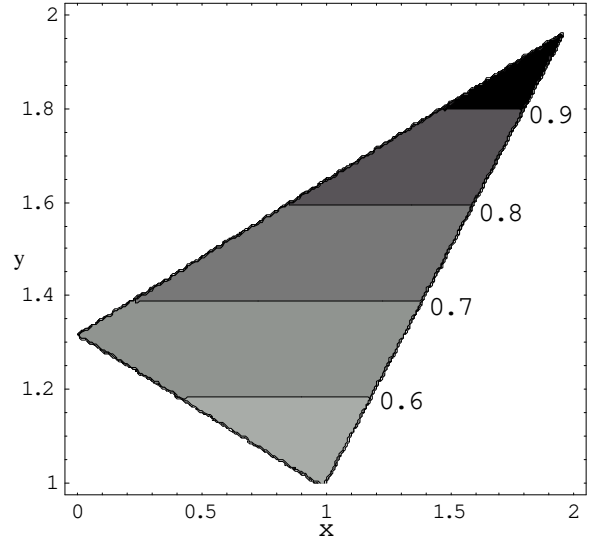}
\caption{Contour plot of $q {\rm Im} \Pi_{s,s'}(q,\omega)$ in units of ($m^2/\pi$)
for $s'=-s=1$
at $B=10^{16}$ G, $T=10^9$ K
and density $n=0.155 \ {\rm fm}^{-3}$ with $x=\beta q$ and $y=\beta \omega$.}
\label{poln}
\end{center}
\end{figure}
%
As is illustrated in
Fig. \ref{poln} for small $T$
the region in the $q$,$\omega$-plane where $q {\rm Im}  \Pi (\vec{q},\omega)$
is finite, is essentially bounded by three straight lines (which become exact
limits for $T \rightarrow 0$); these boundaries are essentially  determined
by the fact that $q < \omega$
and the condition  $p_{\rm{min}} \le  p^{-}_{Fn}$ .
 The latter can also be
expressed (neglecting all terms in $p_{\rm{min}}$
except the leading order terms in $m$) as
$\omega \ge h^-, \omega \le h^+$ with
 $h^{\pm} \approx (2m |\mu_n| B \pm p^{-}_{Fn}  q)/ (m^2+ (p^{-}_{Fn})^2)^{1/2}$.
%
\subsection{Emissivity}
We have only to consider the case $s' = - s = 1,$
in which case $M_{ss'}^2= G_F^2c^2_A/{2}$.
By integrating over the neutrino momenta in Eq. (\ref{emiss})
and introducing the dimensionless parameters
 $y=\beta \omega$ and $x=\beta q$, the
emissivity can be expressed as
\begin{eqnarray}
\epsilon_{\nu \overline{\nu}}= \frac{G_F^2 c^2_A m^2}{2 (2
\pi)^5} T^7
\int_{0}^{\infty} dy \frac{y^4}{e^{y}-1} \nonumber\\
\int_{0}^{y} dx
\Bigg( \ln  \frac{e^{-\beta E_{\rm{min}}}+1}
{e^{-\beta E_{\rm{min}} -y}+1}
-  \ln  \frac{e^{-\beta E_{\rm{max}}}+1}
{e^{-\beta E_{\rm{max}} -y}+1} \Bigg).
\label{emissnpbs}
\end{eqnarray}
To obtain some  insight in the dependence of the emissivity
on $B$  we  distinguish three different regions of $B$ (for simplicity
we take
$|q| = \omega$, so that the integral over $x$
can be replaced by $2y/5$)
\begin{itemize}
\item For weak magnetic
fields ($|\mu_n| B \ll T$)
the region in which $E_{min} < 0 $
 is proportional to $ \mu_{n} B/T$ and peaks around
 $y \approx \mu_{n} B/T$.
For $y \ll 1$
one obtains
\begin{eqnarray}
\epsilon_{\nu \overline{\nu}} \approx
\frac{G_F^2 c^2_A m^2}{30 (2
\pi)^5} T^7  (y_{max}^6-y_{min}^6),
\end{eqnarray}
where $y_{max/min}= \left(2\mu_nB/mT\right)
(\sqrt{m^2+ (p^{-}_{Fn})^2} \pm p^{-}_{Fn}).$
Hence for fixed $T$, one finds
\begin{equation}
\epsilon_{\nu \overline{\nu}} \propto B^6.
\end{equation}
\item For strong magnetic fields ($ |\mu_n| B \sim T$)
the Bose function in Eq. (\ref{emissnpbs}) must be kept and
the emissivity becomes
\begin{equation}
\epsilon_{\nu \overline{\nu}} \approx \frac{G_F^2 c^2_A m^2}{5 (2
\pi)^5} T^7
\int_{y_{min}}^{y_{max}} dy
\frac{y^6}{e^{y}-1}.
\end{equation}
The integral peaks around $|\mu_n| B \approx 3 T$ with a width  $6T (p^-_{Fn}/m)$.
As discussed in section 7 for $B$ values in the range  $|\mu_n| B \sim T$
the emissivity becomes comparable to conventional (modified Urca)
process.
\item From Eq. (\ref{emissnpbs}) we see that for superstrong magnetic fields
($ |\mu_n| B \gg T $)
the emissivity falls off  exponentially with the magnetic field
for fixed $T$.
\end{itemize}
\section{Neutrino-pair bremsstrahlung from protons \
($p \rightarrow p + \nu + \overline{\nu}$)}
In a magnetized matter
in addition to the spin-magnetic field
 interaction the charged particles (protons, electrons)
 are grouped into Landau levels.
The proton's Fermi momentum
in a magnetic field for given Landau level number $N$ and spin $s$ is
given by (see Appendix A)
\begin{displaymath}
p^{s,N}_{Fp}=(E^2_{Fp}-m^2-(2N+1-s g_p)eB)^{1/2},
\end{displaymath}
where $g_p$ is the proton $g$-factor. \\
The population of Landau levels leads to a smearing
of the transverse momentum, $\Delta p_\perp \sim \sqrt{eB}, $ and
as a consequence
the condition for energy/momentum conservation
is softened.
In the special case of a superstrong magnetic field,
$eB > (p^{+0}_{Fp})^2$,
only the lowest ($N=0$) Landau level is occupied.
\subsection{Emissivity}
In the present case it is convenient to define a reduced
proton polarization function for a specific Landau level
$N= p^2_{\perp} /(2eB)$
\begin{eqnarray}
{\rm Im} \Pi_{N',N,s',s} (q_z,\omega_z)=\int dp_{z}
[f(E_{pz,N,s})- \nonumber\\
f(E_{pz',N',s'})] \,
\delta(E_{pz,N,s}-E_{pz',N',s'}+\omega_z+U_{N',N}),
\label{pol'}
\end{eqnarray}
where $\omega_z=(\omega^2-\omega_{\perp}^2)^{1/2}$ with
$\omega_{\perp}=(q^2_x+q^2_y)^{1/2}$,
$U_{N',N}$ is the energy difference between the Landau levels $N,N'$
and the proton energies $E_{pz,N,s}$ are defined in  Appendix A.
In evaluating Eq. (\ref{pol'}) the integral over $p_z$ is replaced by
$(m/p^{s,N}_{Fp})
\int dE_{pz,N,s},$ and  after integration over $E_{pz,N,s}$
one obtains
\begin{eqnarray}
{\rm Im} \Pi_{N',N,s',s} (q_z,\omega_z)=\frac{m}{p^{s,N}_{Fp}}
[f(\tilde{E}_{a})+f(\tilde{E}_{b}) \nonumber\\
-f(\tilde{E}_{a}+\omega_z+U_{N'N})
-f(\tilde{E}_{b}+\omega_z+U_{N'N})]
\end{eqnarray}
with
\begin{eqnarray}
\tilde{E}_{a,b}= (m^2 + \tilde{p}^2_{a,b}+
(2N+1- s g_p) eB)^{1/2},
\end{eqnarray}
where $\tilde{p}_{a}$ and $\tilde{p}_{b}$
follow from the condition of conservation
of $z$-momentum and energy. To obtain real solutions
for $\tilde{p}_{a}$ and $\tilde{p}_{b}$   the conditions
$d \equiv (q^2_z - \omega^2_z + 4 m \mu_p B)^2-4 m^2
(\omega^2_z-q^2_z)>0$ (neglecting some small terms) and $s=-s'=1$
should be satisfied. The result for $\tilde{p}_{a}$ and $\tilde{p}_{b}$ is
\begin{eqnarray}
\tilde{p}_{a,b}(q_z,\omega_z)=
\frac{|-(4 m \mu_p B + q^2_z - \omega^2_z) q_z
\pm \omega_z  \sqrt{d}|}{2(\omega^2_z-q^2_z)}.
\end{eqnarray}
We note that the polarization function
essentially vanishes unless
$\tilde{p}_{a}$ and $\tilde{p}_{b}$
are close to the Fermi momentum $p^{s,N}_{Fp}=
\sqrt{(p_{Fp}^{s,0})^2-2NeB} $.
In the case of a superstrong magnetic field only the $N'=N=0$ states
are occupied and the contribution to
${\rm Im} \Pi_{N',N,s',s} (q_z,\omega_z)$ comes
only from the lines in the $\omega_z$,$q_z$-plane defined by
(neglecting all terms in $p_{{a}}$
except the leading order terms in $m$)
\begin{eqnarray}
\omega_z^{\pm} \approx \frac{2m \mu_p B \pm q_z p_{Fp}^{+0}}
{\sqrt{m^2+2 (1-g_p) m \mu_B B+
(p_{Fp}^{+0})^2}}.
\end{eqnarray}
For weaker $B$-fields a larger space in the $q_z$,$\omega_z$-plane
contributes eventually leading to a situation
similar to that for neutrino-pair bremsstrahlung from neutrons. \\
The emissivity can be written as
\begin{eqnarray}
\epsilon_{\nu \overline{\nu}}= \sum_{N',N,s',s}^{}
\frac {6}{4(2 \pi)^4} \int_{}^{} d\omega
\, dq_z \, d\omega_z \, g_B(\omega)  \,
\frac{\omega^4}{6}  \nonumber\\
{\rm Im} \Pi_{N',N,s',s} (q_z,\omega_z)
\, |M_{N',N,s',s}|^2,
\label{emis'}
\end{eqnarray}
where
\begin{displaymath}
|M_{N',N,s',s}|^2=
\sum_{R' R}  |{\cal M}_{R',R,N',N,s',s}|^2 =
 \frac{eB}{2 \pi}
\end{displaymath}
\begin{displaymath}
\frac{G_F^2}{4}
[ c_V^2 \delta_{s',s} +2c_A^2 \delta_{s',-s}
+c_A^2 \delta_{s',s}] I^2_{N',N}(v)
\end{displaymath}
with $v=(\omega^2-\omega^2_z)/(2eB)$, $R$ and $R'$ the
guiding center quantum numbers
and $I_{N,N'}$
the associated Laguerre polynomials (see
Appendix A).
As in the  case of neutron bremsstrahlung only one spin
configuration contributes ($s=-s'=1$).
For strong magnetic fields, $B \gg T^2/(4 m \mu_B)$,
the dominant contribution to the emissivity comes from
$N'=N$ in which case
the weak interaction matrix can be simplified in case of $Nv\ll 1$
\begin{eqnarray}
\sum_{R',R }  |{\cal{M}}^2_{R',R,N,N,-,+}|^2 =
\frac{G_F^2}{2} c_A^2  \frac{eB}{2 \pi}
I^2_{N,N}(v)   \nonumber\\
\approx  \frac{G_F^2}{2} c_A^2 \frac{eB}{2 \pi}.
\label{Lag}
\end{eqnarray}
Using the fact that $U_{N'N}$ vanishes for $N'=N$,
in this case the emissivity becomes
\begin{eqnarray}
\epsilon_{\nu \overline{\nu}}= \sum_{N=0}^{N_{max}}
\frac{24 m c^2_A G_F^2  \mu_B B T^7} {4 (2 \pi)^5 }
\int_{}^{} dy \frac{1}{e^{y}-1}
\frac{y^4}{6}
\nonumber\\
\int_{0}^{y} dy_z   \int_{0}^{y_z} dx_z \, {\rm Im} \Pi_{N,N,-,+} (x_z,y_z),
\label{superstrong}
\end{eqnarray}
where  $y=\omega/T$, $y_z=\omega_z/T$, $x_z=q_z/T$ and
$N_{max}=(p^{+0}_{Fp})/(4 m \mu_B B)$.
In weak magnetic fields when eq. (\ref{Lag})
is not a good approximation, the summation over $N$ must be restricted
to $(4 m \mu_B B)/(36 T^2)$.
In general the integral over $x_z$ must be carried out numerically.
In case of $\mu_p B \approx T$ the main contribution comes from
$y_z= y$, and
as a result the emissivity of neutrino-pair bremsstrahlung from protons
is very similar
to neutrino-pair bremsstrahlung from neutrons.
On the other hand for weaker fields
the emissivity of neutrino-pair bremsstrahlung from protons is larger due
to the smearing of the transverse momenta of the protons.
\section{Direct Urca process \ ($n \rightarrow p + e + \overline{\nu}_e$; $p + e
\rightarrow n + \nu_e$) }
It is well known that
the direct Urca process can occur only if
the Fermi momenta satisfy the inequality
\be\label{urcatriang}
p_{Fp}+p_{Fe}>p_{Fn}
\ee
(except for a small thermal smearing),
i.e.  the proton concentration $x_p= {n_p}/(n_p +n_n)$
needs to be larger than 1/9 (Lattimer et al. \cite{Lat2}).
Recently it was shown (Leinson \& Perez \cite{Leinson};
Baiko \& Yakovlev \cite{Baiko}) that the presence of a magnetic field
has an effect on the Urca process;
in particular it  leads to a non-vanishing emission rate
in case the triangular condition (\ref{urcatriang})
is not satisfied (the so-called classically forbidden region).
This is caused by  a smearing of the momenta
in the presence of a magnetic
field which leads to a softening of the sharp border between the allowed
and forbidden regimes.
Since the emissivities due to the Urca process in superstrong
$B$-fields  (Leinson \& Perez \cite{Leinson}) and
in arbitrary  $B$-fields (Baiko \& Yakovlev \cite{Baiko})
have been derived previously,
we provide below a brief discussion
for the purpose of  completeness.
\subsection{Emissivity}
The emissivity is calculated using the quasi-particle approximation
for the polarization
function where the electrons and protons occupy the
 Landau levels $N_e$, $N_p$,
respectively,
\begin{eqnarray}
\epsilon_{\nu \overline{\nu}}= 2 \sum_{ N_e N_p s_n s_p  } \frac{ 1}{
(2 \pi)^6}
\int d^3p_n \, dp_{ez}
d^3p_{\nu} \, dp_{pz} \nonumber\\
\delta(p_{nz}-p_{pz}-p_{ez}-p_{{\nu}z})
 \sum_{R_e R_p}|{\cal M}_{R_{e},R_{p},N_{e},N_{p},s_{n},s_{p}}|^2
\nonumber\\ \omega_{\nu} f(E^n)  (1-f(E^p)) (1-f(E^e))
\nonumber\\ \delta
(E^n+V_{np}-E^p-E^e+\omega_{\nu}),
\label{emissivity}
\end{eqnarray}
where $V_{np}=E_{Fp}+E_{Fe}-E_{Fn}$ and
the neutron, proton, and electron energies
are given in Appendix A and the matrix element ${\cal M}$
in Appendix B. The factor 2 comes from the contribution of the
inverse reaction.
\subsubsection{Superstrong magnetic fields}
First we consider superstrong magnetic fields where the
electrons and protons populate the ground state Landau levels
($N_e=N_p=0$)
with spin parallel to $B$,
so that $|p_{pz}| \approx p^{+0}_{Fp} $ and $|p_{ez}| \approx p^0_{Fe}$.
Neglecting the neutrino momentum the matrix element
in Eq. (\ref{emissivity}) simplifies to
 $\sum_{R_e R_p} |{\cal M}|^2= C_{s_n} \exp(-
p^2_{n \perp}/2eB)$ with $C_{s_n}=G_F^2 [(1-2 c_A + c^2_A)
\delta_{s_n,1}+4 c^2_A
\delta_{s_n,-1}]/2,$
and
$p^2_{n \perp}=
(p^{s_n}_{Fn})^2-(p_{pz} +
p_{ez})^2$.
Evaluating the energy integrals leads to
\begin{eqnarray}
\epsilon_{\nu \overline{\nu}}= \sum_{s_n} C_{s_n} \frac{4 (120+6 \pi^2) B
e}{ (2
\pi)^5}
\frac{m_n m_p}{p^{+0}_{Fp}} T^6 \nonumber\\
(\Theta(p^{s_n}_{Fn})
\exp[- (p^{s_n}_{Fn})^2/2eB]
+ \Theta(p^{s_n}_{Fn}-|
p^{+0}_{Fp}+p^0_{Fe}|) \nonumber \\
\exp[-((p^{s_n}_{Fn})^2-(p^{+0}_{Fp}+p^0_{Fe})^2)/2eB]).
\label{emisground}
\end{eqnarray}
The $\Theta$ functions
correspond to conservation of momentum in the $z$-direction;
it is worth noting that the triangular condition expressing
the momentum conservation in the presence of
a superstrong $B$-field, $|p^{+0}_{Fp}+p^0_{Fe}|<|p^{s_n}_{Fn}|$,
is the opposite of the one found for the $B$=0 case. \\
The expression (\ref{emisground}) for the emissivity
has been obtained previously by Leinson \& Perez (\cite{Leinson})
and Baiko \& Yakovlev (\cite{Baiko}) .
The factor $C_+=G_F^2 (1-c_A)^2/2$ in Eq. (\ref{emisground})
agrees with the result of Baiko \& Yakovlev (\cite{Baiko}),
but differs from the one given by Leinson \& Perez (\cite{Leinson}).
\subsubsection{Weak magnetic fields}
In the case of weak magnetic fields, i.e.  $eB<p^2_{Fp}$,
the summation over the Landau levels in Eq. (\ref{emissivity})
is replaced by an integral\footnote{In this Sect. we use the
definition of the Fermi momenta as
in the field free case.}. Also the  $I_{NN}$ functions
in the matrix element given by Eq. (\ref{urcama}) of Appendix B
are replaced by their small-$B$ asymptotes
\begin{eqnarray}
 2eB \sum_{N_e N_p} I^2_{N_e N_p}(v) \rightarrow \nonumber\\
\frac{1}{2Be} \int
dp^2_{e \perp} dp^2_{p \perp} A(p_{p \perp},p_{e \perp},B) \, Ai^2(z)
\label{inte}
\end{eqnarray}
with $Ai(z)$ being the Airy function with
\begin{displaymath}
z=\frac{[p^2_{n \perp}-(p_{p \perp}+p_{e \perp})^2] (p_{p \perp}+p_{e \perp})^{1/3}}
{(2eB)^{2/3} (p_{p \perp}+ p_{e \perp})^{4/3}},
\end{displaymath}
$v=p^2_{n \perp}/2eB$, and
\begin{displaymath}
A(p_{p \perp},p_{e \perp},B)= \frac{(2eB)^{2/3}}{(p_{p \perp} + p_{e \perp})^{2/3}
(p_{p \perp} p_{e \perp})^{1/3}}.
\end{displaymath}

Neglecting the neutrino momentum in the
$z$-direction in the delta
function in Eq. (\ref{emissivity})
in comparison with the momenta in the $z$-direction
of the other particles and performing
the integral over $d\cos( \theta_n )$,
 the emissivity can be written as
 (Baiko \& Yakovlev \cite{Baiko})
\begin{eqnarray}
\epsilon_{\nu \overline{\nu}} = \frac{8(1+3 c^2_A) G_F^2 (120+6 \pi^2) m_n m_p
p_{Fp} p^2_{Fe}}{(2 \pi)^5 eB} \nonumber\\
\int d\cos( \theta_p ) \,
d\cos( \theta_e) \,  A(p_{p \perp},p_{e \perp},B) \, Ai^2(z) \nonumber\\
\delta(p_{Fn}-|p_{Fp} \cos( \theta_p) +
p_{Fe} \cos( \theta_e)|).
\end{eqnarray}
In the classically forbidden domain  for the direct Urca process
for finite $B$ the reaction becomes possible due to the tunnelling mechanism.
In the quasi-classical approach the emissivity can be expressed as
(Baiko \& Yakovlev \cite{Baiko})
 \be
 \epsilon_{\nu
\overline{\nu}}^{forbidden}(B \ne 0)
=R(x,y) \ \epsilon_{\nu \overline{\nu}}^{allowed}(B=0)
\ee
with
\begin{eqnarray}
R(x,y)\approx \sqrt{\frac{y}{x+12y}} \,  \frac{3}{x^{3/2}}
\, \exp(\frac{-x^{3/2}}{3}),
\end{eqnarray}
where $y=N_{Fp}^{2/3}= (p_{Fp}^2/(2eB))^{2/3}$ and $x$ is a measure of the violation
of the triangular condition,
\begin{eqnarray}
x=\frac{p^2_{Fp}-(p_{Fp} + p_{Fe})^2}{p^2_{Fp}
N^{-2/3}_{Fp}}. \
\end{eqnarray}
One sees that with increasing $x$,
the emissivity decreases exponentially,
and therefore the effect is important only if $p_{Fn}$ does not exceed
$(p_{Fp}+p_{Fe})$ significantly.  \\
On the other hand in the classically allowed region, where the
inequality is satisfied, the $B$-field will give rise to
small quantum oscillations of the emissivity.
\section{Equation of state (EoS)}
In order to estimate the effect of a magnetic field on the
various cooling processes in a neutron star and to compare
with the conventional result  we employ a simple model EoS.
It is assumed that the neutron star matter consists of neutrons,
protons and electrons only ($npe$-matter) with two conditions imposed:
charge neutrality, $n_p=n_e$, and $\beta$-equilibrium,
$\phi_n = \phi_p + \phi_e$\footnote{$\phi_{i}$ is the
chemical potential of a particle; it can be obtained
from the relation \ $\phi_i=dE/dn_i$.}.
The EoS consists of
the energy density as a function of the densities of the particles.
The non-relativistic energy density can be decomposed as
(Lattimer \& Swesty \cite{Lattimer};
Balberg \& Gal \cite{Balberg})
\begin{eqnarray}
E=E_{kin}+E_{mag}+E_{mass}+E_{pot}+E_{lep}.
\end{eqnarray}
Here the kinetic energy density is the sum
of the neutron
\begin{eqnarray}
E^n_{kin}=\frac{3}{5} \sum_{s_n} \frac{(p^{s_n}_{Fn})^2}{2 m_n} n^{s_n}_n
\end{eqnarray}
and proton contribution
\begin{eqnarray}
E^p_{kin}=\frac{1}{3}
\sum_{s_p}
 \frac{(p^{s_p,0}_{Fp})^2}{2 m_p} n^{s_p}_p,
\end{eqnarray}
where the proton density for spin $s_p$ is
\be
 n^{s_p}_p= \sum_N \frac{eB}{2\pi^2} \sqrt{(p^{s_p,0}_{Fp})^2-2NeB},
\label{denpro}
\ee
because for finite $B$ the integrals over the transverse momentum of
the proton with
respect to the magnetic field can be replaced by a summation
over the Landau levels.
The interaction of the magnetic field with the spin of the neutron and the
proton is
\begin{eqnarray}
E_{mag}=-\mu_n B \sum_{s_n} s_n  n^{s_n}_n -
\mu_p B \sum_{s_p}
s_p n^{s_p}_p.
\end{eqnarray}
The $E_{mass}$ term contains the masses of the  two nucleons
\begin{eqnarray}
E_{mass}= m_n \sum_{s_n} n^{s_n}_n + m_p \sum_{s_p}
n^{s_p}_p.
\end{eqnarray}
The $E_{pot}$ term is the potential energy density, which is
parametrized as (Lattimer \& Swesty \cite{Lattimer})
\begin{eqnarray}
E_{pot}=a n^2 + b n^{1+d} + \sum_{s_n,s_p} 4 c n^{s_n}_n n^{s_p}_p,
\label{Epot}
\end{eqnarray}
where $n=n_p+n_n$ is the total nucleon density.  The
last term on the rhs of Eq. (\ref{Epot}), which corresponds to the
symmetry energy, influences to a large extent the proton fraction
$n_p/(n_p+n_n)$.
Finally the energy density of the electrons is
\begin{eqnarray}
E_{lep}=E_{Fe} n_{e},
\end{eqnarray}
where
\begin{eqnarray}
n_e
=\sum_{N_e} \frac{g_{N_e}eB}{2 \pi^2} \sqrt{(p_{Fe}^{0})^2-2 N_e e B}
\label{denelec}
\end{eqnarray}
with $g_{N_e}=1$ for $N_e=0$, $g_{N_e}=2$ for $N_e > 0$ and
$N_e$ is limited from above by $N_{e}^{max}=(p^0_{Fe})^2/(2eB)$.
\\
We obtain the EoS of the $npe$-matter in magnetic fields in
the standard manner, by assuming a given nucleon density  $n$ and $T=0$
and solving the
equations of charge neutrality and $\beta$-equilibrium.
In two limiting cases the solutions
are straightforward. \\
i) $B=0$.
The summations
over the proton and electron concentration in Eqs. (\ref{denpro}) and
(\ref{denelec}) can be replaced by integrations
and the densities $n_e$, $n_p$ and $n_n$ can be calculated
in the standard way using $n_i=p^3_{Fi}/(3 \pi^2)$.
Using the $\beta$-equilibrium condition, $\phi_e=\phi_n-\phi_p$,
the chemical potential of the electron is given by
\begin{eqnarray}
\phi_e=\frac{(3 \pi^2 n_{n})^{2/3}}{2 m_n}  - \frac{(3 \pi^2 n_{p})^{2/3}}{2
m_p}  + 4 c (n_p-n_n) + \Delta
\end{eqnarray}
with $\Delta$ the mass difference between neutron and proton. \\
ii) Superstrong magnetic fields,
\label{Blim2}
 $(p^{+0}_{Fp})^2 <(2e B)$, in which case
the protons and electrons are in the ground state Landau level.
Then the electron chemical potential is
\begin{eqnarray}
\phi_e=\frac{1}{2} \sum_{s_n} \frac{(6 \pi^2 n^{s_n}_n)^{2/3}}{2 m_n}  -
\frac{1}{2 m_p} \Big(\frac{2 \pi^2
n_{p}}{e B} \Big)^2 \nonumber\\
+ \frac{g_p-1}{2 m_p} \, e B  + 4 c (n_p-n_n) + \Delta.
\end{eqnarray}
In this case the proton fraction depends on  the magnetic field.
In Eq. (\ref{Epot}) we take   $d=2$, $a=-285.1 \
{\rm MeV} \ {\rm fm}^{3}$, $b=968 \ {\rm MeV} \ {\rm fm}^{6}$, $c=-107.1 \
{\rm MeV} \ {\rm fm}^{3}$ (Lattimer \& Swesty \cite{Lattimer}). \\
\begin{figure}
\begin{center}
\includegraphics[angle=0,width=0.45\textwidth]{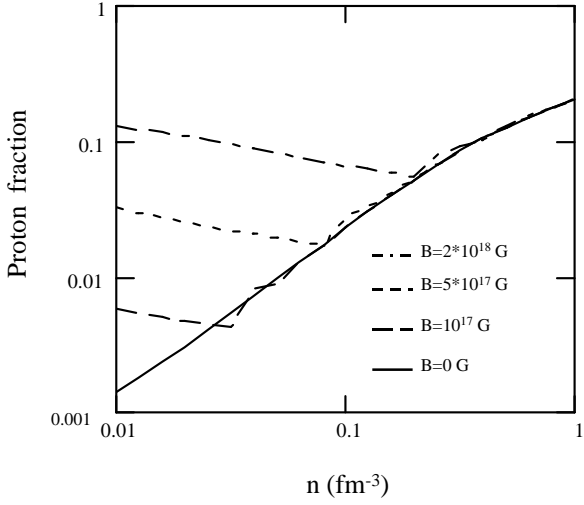}
\caption{Proton fraction as a function of density
for various B-fields}
\label{fraction2}
\end{center}
\end{figure}
In Fig. \ref{fraction2} the resulting proton fraction in the absence
 and in the presence of magnetic fields is plotted.
The kinks in the proton fraction at certain densities correspond
to the occupation of a next Landau level (see also Suh \& Mathews \cite{Suh}).
For superstrong magnetic fields these kinks strongly influence the
proton fraction, but
for weaker magnetic fields ($B \leq 10^{17} G$ and $n \geq n_0$
where   $n_0= 0.155\ {\rm fm}^{-3}$ is the saturation density) they
do not affect the proton fraction.
Our results are in agreement with the previous results
for the proton fraction derived from different equations of
state (Lai \& Shapiro \cite{Lai};
Suh \& Mathews \cite{Suh}, Broderick, Prakash
and Lattimer \cite{Bro}).
\section{Results}
\begin{figure}
\begin{center}
\includegraphics[angle=0,width=0.45\textwidth]{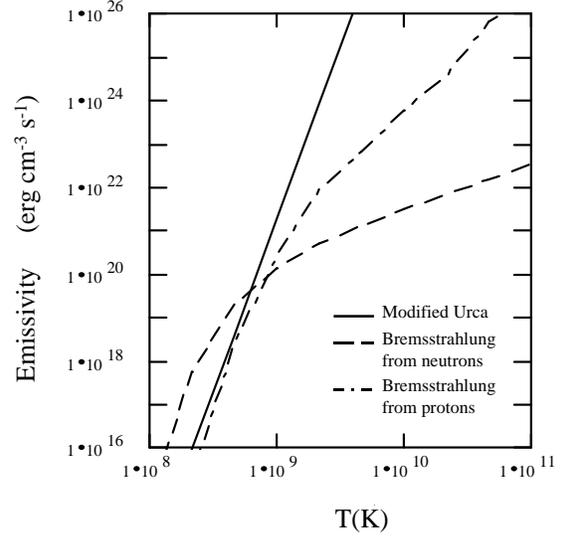}
\caption{Emissivities of various processes
at saturation density $n=n_0$ and for $B=10^{16} {\rm G}$.
The contribution of the direct Urca process is negligible.}
\label{simulnb16}
\end{center}
\end{figure}

\begin{figure}
\begin{center}
\includegraphics[angle=0,width=0.45\textwidth]{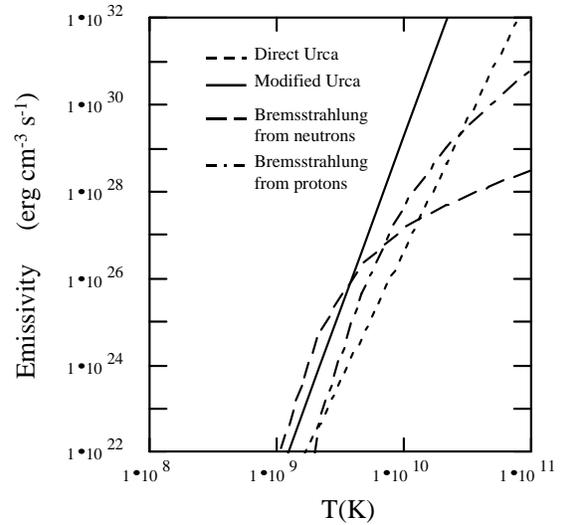}
\caption{Same as in Fig. 3 but for $B=10^{17} {\rm G}$.}
\label{simulnb17}
\end{center}
\end{figure}

\begin{figure}
\begin{center}
\includegraphics[angle=0,width=0.45\textwidth]{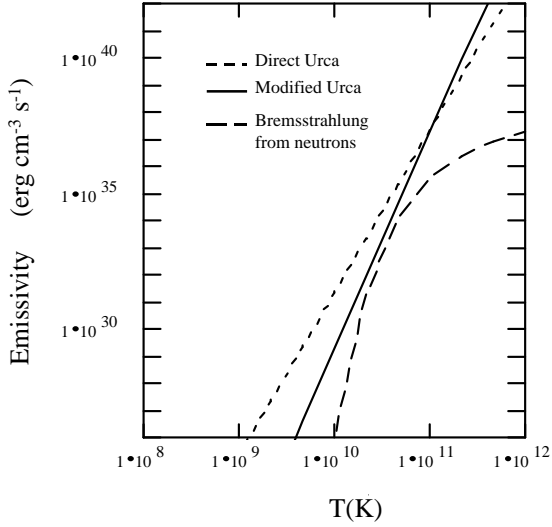}
\caption{Same as in Fig. 2 but $B=2\times 10^{18} {\rm G}$}
\label{simulnb19}
\end{center}
\end{figure}

The emissivities for the various
energy loss processes are compared  in Figs. \ref{simulnb16},
 \ref{simulnb17} and \ref{simulnb19}
for $n=n_0$ for three different magnetic field strengths
 ($10^{16}$,$10^{17}$ and $2\times  10^{18}$ G, respectively).
In the first two values of the $B$-field the EoS based on $B=0$
is used (because in these cases the influence of the magnetic field
on the EoS itself is small); for the third value of the $B$-field
we use the EoS for superstrong magnetic fields.
\\ To enable a comparison with the neutrino processes which
are routinely included in the cooling simulations of
neutron stars, we show also the emissivity
due to the modified Urca process in the zero field limit.
The relevant expression for the
emissivity of the modified Urca process, without corrections for the
magnetic field,  and valid for $m^* =m$ is (Friman \& Maxwell \cite{FM})
\begin{eqnarray}
\epsilon_{\nu \overline{\nu}}= 1.8\times 10^{21} (n/n_0)^{2/3}T_9^8
\ {\rm ergs} \ {\rm cm}^{-3} \ {\rm s}^{-1}
\end{eqnarray}
with $T_9$ the temperature in units of $10^9$ K.

The temperature region where the one-body neutrino-pair bremsstrahlung
is important increases with increasing magnetic field
(Figs. \ref{simulnb16} and  \ref{simulnb17}).  The pair bremsstrahlung
from neutrons  is efficient whenever $|\mu_n| B \sim T$, since then
the energy involved in the spin-flip is of the same order of magnitude
as the thermal smearing of the Fermi surface.
The temperature at which neutrino-pair bremsstrahlung from neutrons
becomes comparable to the competing processes
roughly coincides with this condition. For lower temperatures the
emissivity drops exponentially, because the energy transfer becomes larger
than the thermal smearing.
Neutrino-pair bremsstrahlung from protons
is important when  $\mu_p B \sim T$.
The emissivity due to the protons increases faster than
the emissivity due to the neutrons with the temperature, as
the smearing of the proton transverse momenta provides
an additional relaxation on the kinematical constrains.
However for temperatures smaller than $\mu_p B \sim T$ the
emissivity drops just as for neutrons exponentially.
\\ As seen from Figs. \ref{simulnb16} and \ref{simulnb17} the emissivity
of the modified Urca process is
larger than that of neutrino-pair bremsstrahlung
from neutrons and protons at high temperatures mainly
due to the different temperature  dependencies of these processes
($\propto T^7$ for the one-body bremsstrahlung as compared to
$\propto T^8$ for the modified Urca).
In the case of a superstrong magnetic field the large uncertainty in
the transverse momenta of the protons and electrons allows
the direct Urca process to occur (see Fig. \ref{simulnb19}) and
its emissivity dominates the emissivity of any other processes.
\begin{figure}
\begin{center}
\includegraphics[angle=0,width=0.45\textwidth]{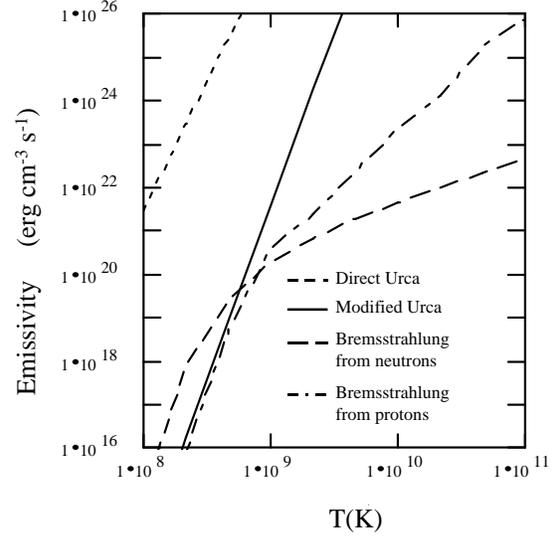}
\caption{Emissivity at $n=3 n_0$ and $B=10^{16} {\rm G}$.}
\label{simul3n}
\end{center}
\end{figure}
In Fig. \ref{simul3n} the emissivities are shown for $n=3 n_0$ and
$B=10^{16}$ G.
At this density the triangle condition
$p_{Fp}+p_{Fe} > p_{Fn}$ is satisfied, so that the direct Urca process
is allowed and dominates the cooling. The emissivities of the other
processes are slightly larger than those shown  in Fig. \ref{simulnb16}
due to the fact that the density is larger.
\section{Conclusion}
  We have studied the neutrino emissivity of strongly magnetized
   neutron stars due to the one-body processes driven  by the
   charged and neutral current couplings between the neutrinos
   and baryons. We have shown that, in addition to the well-known charged
   current process (the direct Urca reaction),  there is a new
   channel of energy loss - the one-body neutrino pair-bremsstrahlung
   in a magnetic field. The process  does not have an analogue in the
   zero field limit and competes with the modified (two-body) bremsstrahlung
   process as the dominant neutral current
   reaction for fields on the
   order $10^{16}-10^{17}$ G and temperatures a few times $10^9$ K.
   For superstrong magnetic fields in
   excess of $10^{18}$ G the direct Urca process takes over. \\
   Our numerical evaluation of the emissivities of several competing
   reactions, which is  based on a simple parametrization
   of the EOS of $npe$-matter in a strong magnetic field, shows that
   under certain conditions
   the emissivities of the one-nucleon processes, such as
   the direct Urca and the one-body bremsstrahlung,
   are of the same order of magnitude or dominate the standard
   processes commonly included in the cooling simulations in
   the zero-field limit.
\appendix
\section{Electrons, protons and neutrons in a magnetic
field}
\label{A}
The wave equation for a fermion with
charge $q$ and mass $m$ can be written as
(Itzykson \& Zuber \cite{Itzykson}):
\begin{equation}
(i \not \hspace{-0.5mm} \partial + q \not \hspace{-1.5mm} A  - (m-\frac{\Delta g}{2}
\mu_B  \sigma^{\mu \nu} F_{\mu \nu } ) )  ,
\Psi (x) = 0  ,
\label{general}
\end{equation}
with $\mu_B=e/2m$, $ \not \hspace{-0.5mm} \partial =\gamma_{\mu} \partial^{\mu}$,
$ \not
\hspace{-1.5mm} A = \gamma_{\mu} A^{\mu}$
and $\Delta g$ the anomalous gyromagnetic factor.
Here
$\gamma^{\mu}$ is a Dirac matrix, $A^{\mu}$ the vector potential, $F_{\mu
\nu}=\partial_{\mu} A_{\nu} - \partial_{\nu} A_{\mu}$  and  $\sigma^{\mu \nu}= \frac{i}{2}
[\gamma^{\mu},\gamma^{\nu}]$.
Or equivalently
\begin{eqnarray}
((i \partial + q A)^2+ \frac{q}{2} \sigma^{\mu \nu} F_{\mu \nu} - \nonumber\\
(m- \frac{\Delta g}{2}
\mu_B \sigma^{\mu \nu} F_{\mu \nu})^2) \Psi (x)
 = 0.
\end{eqnarray}
In particular, if one chooses
the  Landau gauge $A^{\mu}=(0,-\frac{By}{2},\frac{Bx}{2},0)$
with the magnetic field in the $z$-direction, one finds
\begin{equation}
((i \partial + q A)^2+ s q B -
(m- \Delta g  s \mu_B B)^2) \Psi (x) = 0.
\end{equation}
\subsection{Electron}
For an electron  ($q=-e$ and $\Delta g=0$)
the energy eigenvalues
are given by
\be
E_{N,s}=(m^2+p^2_z+2NeB)^{1/2},
\ee
 with  $N$ denoting the Landau level number.
The eigenfunctions are factorized  (Sokolov \& Ternov \cite{ST};
Arras \& Lai \cite{Arras})
\begin{displaymath}
\psi_e = e^{i p_{z} z - i E_{N,s} t}
U_e(\rho,\phi),
\end{displaymath}
\begin{eqnarray}
U_e(\rho,\phi) = \sqrt{\frac{eB}{2 \pi}} e^{i(N - R)
\phi}
\left (\begin{array}{l}
V_a I_{N-1,R}(t)  e^{-i \phi}\\
i V_b I_{N,R}(t) \\
V_c I_{N-1,R}(t)  e^{-i \phi}  \\
i V_d I_{N,R}(t)  \\
\end{array} \right)
\end{eqnarray}
and
\begin{displaymath}
I_{N,R}(t)= \sqrt{\frac{R !}{N !}}
e^{-t/2} t^{(N-R)/2}
\tilde{L}_{R}^{N-R}(t),
\end{displaymath}
where $\tilde{L}_{R}^{M}(t)$
\begin{displaymath}
 = \left\{ \begin{array}{llr}
L_{R}^{M}(t) & \ {\rm if} & \ \ M \geq 0 \\
(-1)^{|M|} t^{|M|} L^{|M|}_{R-|M|}(t)  & \  {\rm if} & \ \ M<0
\end{array} \right.
\end{displaymath}
with $t={\rho}^2eB/2$,  $V_a=C_+ D_+$,
$V_b=\sigma C_- D_+$,
 $V_c=\sigma C_+ D_-$,  $V_d=C_- D_-$,

$$C_\pm = \frac{1}{\sqrt{2}} \left(1 \pm \sigma
\frac{p_{z}}{(p^2_{z}+4 m \mu_B B
N)^{1/2}}\right)^{1/2},$$ \\
$$D_\pm = \frac{1}{\sqrt{2}} \left(1 \pm \frac{m}{E}
\right)^{1/2},$$ and  $\sigma$ is
the longitudinal  spin projection along $\vec{p} +e\vec{A}$;
$L_{R}^{M}(t)$
is the generalized Laguerre function.
The Landau level number $N$ and the guiding center quantum number $R$
are positive integers.
 (Note that in the lowest energy state ($N=0)$ the spin
 $s$ can only have the value -1). \\
In matter in the presence of a magnetic field
the Fermi energy is given by
 $E_F=(m^2+(p^N_{Fe})^2+2 N e B)^{1/2}$
 so that a magnetic field will give
rise to a difference
in the Fermi momenta $p^N_{Fe}$.
The degeneracy is 1 in
case of $N=0$ and 2 in case of $N>0$.
\subsection{Proton}
For a proton ($q=e$ and $\Delta
g=g_p -1$ with $g_p=2.79$)
the energy eigenvalues
(neglecting
the $B^2$-term)
are
\be
E_{N,s}=(m^2
+p^2_z+(2N+1-s g_p) eB)^{1/2}. \label{proton}
\ee
The non-relativistic  eigenfunction of a proton in a
magnetic field is
\begin{displaymath}
\psi_p = e^{i p_{z} z - i E_{N,s} t} U_p(\rho,\phi),
\end{displaymath}
\begin{equation}
U_p(\rho,\phi) = \sqrt{\frac{eB}{2 \pi}} e^{i(R - N)
\phi}
\left (\begin{array}{l}
\delta_{s,1} I_{R,N}(t) \\
\delta_{s,-1} I_{R,N}(t) \\
0  \\
0  \\
\end{array} \right).
\label{protonfunc}
\end{equation}
 (Note that the role of $R$ and $N$ are interchanged compared to
 the electron case.)
 In matter the Fermi energy is given by
$E_F=(m^2+(p^{s,N}_{Fp})^2+(2N+1-s g_p) eB)^{1/2}$.
\subsection{Neutron}
For a neutron ($q=0$ and $\Delta g=g_n$ with
$g_n=-1.91$)
in not extremely large magnetic fields the $B^2$-term can be neglected;
then the energy eigenvalues are
\begin{eqnarray}
 E_s=(m^2+p^2-2 s g_n m \mu_B
B)^{1/2},
\end{eqnarray}
and the eigenfunction are simple plane waves.
In neutron matter in thermal equilibrium
the Fermi energy is given by
$E_F=(m^2+(p^s_{Fn})^2-2 s g_n m \mu_B B)^{1/2}$.
Hence the neutrons with spin up (down) occupy two Fermi
spheres with Fermi momenta related by
$(p^+_{Fn})^2=(p^-_{Fn})^2+4 m \mu_n B$.
\section{The weak interaction matrix elements}
\label{B}
\subsection{Neutrino-neutron interaction (neutral current)}
The  weak
interaction matrix for the neutron bremsstrahlung is
\begin{equation}
M_{s,s'}=\frac{G_F}{2 \sqrt{2}} \int_{V}^{} d \mathbf{r} \overline{\psi}_{n1} \gamma_\mu
(c_V - c_A \gamma_5)
\psi_{n2} \overline{\psi}_\nu \gamma^\mu (1 -  \gamma_5) \psi_{\overline{\nu}}.
\end{equation}
Here $c_V$ and $c_A$ are vector and axial-vector coupling constants.
The interaction matrix consists of a hadronic part and a leptonic part
\begin{equation}
|M_{s,s'}|^2=\frac{G_F^2}{8} X_{\mu \nu} L^{\mu \nu}
\label{hadron}
\end{equation}
with the leptonic part given by
\begin{displaymath}
L^{\mu \nu} = \frac{2}{pp'}[p^{\mu} p'^{\nu}-g^{\mu \nu} (p \cdot
p') + p^{\nu} p'^{\mu}+ i \epsilon^{\alpha \mu \beta \nu}
p_{\alpha} p'_{\beta}]\nonumber ,
\end{displaymath}
and the hadronic part by
\begin{eqnarray}
X_{\mu \nu}=
 \left (\begin{array}{cccc}
c^2_V \delta_{s,s'} & 0 & 0 & - s'c_V c_A \delta_{s,s'} \\
0 &  c^2_A \delta_{s,-s'} & i c^2_A s \delta_{s,-s'} & 0 \\
0 & -i c^2_A s \delta_{s,-s'} & c^2_A \delta_{s,-s'} & 0 \\
- s'c_V c_A \delta_{s,s'} & 0 & 0 & c^2_A \delta_{s,s'}
\end{array} \right)\nonumber .
\end{eqnarray}
Contracting the hadronic and leptonic tensors and
neglecting terms, which after integration over the neutrino momenta vanish
one finds
\begin{eqnarray}
|M_{s,s'}|^2=\frac{G_F^2}{4} [(c^2_V + c^2_A) \delta_{s,s'}+ 2 c^2_A \delta_{s,-s'}]
\end{eqnarray}
with $c_V=1$ and $c_A=1.26$.
\subsection{Neutrino-proton interaction (neutral current)}
In case of protons the integration
over space coordinates requires special attention,
because the wave functions of protons
are not simple plane waves (Eq. \ref{protonfunc}).
Defining $\vec{\omega}_{\perp} ={(p_{{\nu}x}+p_{\overline{\nu}x})} \hat{e}_{x} +
{(p_{{\nu}y}+p_{\overline{\nu}y})} \hat{e}_{y}$,
 $\vec{x}_{\perp} = x \hat{e}_{x}+y
\hat{e}_{y}$, $v=\omega^2_{\perp}/2eB$ and
carrying out the summations over $R'$ and $R$, leads to
\begin{eqnarray}
 \sum_{R',R} |{\cal M}_{R',R,N',N,s',s}|^2=
  \sum_{R',R} \frac{eB}{2 \pi} |M_{s',s}|^2
 \nonumber\\ \bigg| \int_{0}^{\infty} \rho
d\rho I_{R',N'}(t)  I_{R,N}(t)  \nonumber\\
\int_{0}^{2 \pi} d\phi e^{-i(R'-N') \phi} e^{i(R-N)
\phi}
e^{-i \vec{\omega}_{\perp} \vec{x}_{\perp}} \bigg|^2 \nonumber\\
=  |M_{s',s}|^2 \frac{eB}{2 \pi} I^2_{N',N}(v)
\label{npne}
\end{eqnarray}
with $c_V=-0.08$ and $c_A=-1.26$.
\subsection{Direct Urca process (charged current)}
The interaction matrix is
\begin{eqnarray}
{\cal M}_{R_{e},R_{p},N_{e},N_{p},s_{n},s_{p}}
=\frac{G_F}{\sqrt{2}} \int_{V}^{} d \mathbf{r} \overline{\psi}_{p} \gamma_\mu
(c_V - c_A \gamma_5)
\psi_{n} \nonumber\\
\overline{\psi}_e \gamma^\mu (1 -  \gamma_5) \psi_{\overline{\nu}}.
\end{eqnarray}
The eigenfunctions of the electron, proton and neutron are given in
Appendix A. The function $I_{R_p,N_p}(t)$ with $t=\rho^2 e
B/2$ can be substituted  in the leptonic
part. The hadronic part turns out the same
as for the  neutron bremsstrahlung (see
Eq. \ref{hadron}).
The space integrals can be calculated in the same way as in Eq.
(\ref{npne}) with the result
\begin{eqnarray}
L_{\mu \nu}=
\frac{1}{2p} \sum_{s_e= \pm 1} {\rm Tr}[Y_e  \gamma_{\mu} (1-
\gamma_5) \not \hspace{-0.5mm} p \gamma_{\nu} (1-\gamma_5) \overline{Y_e}] \nonumber\\
\end{eqnarray}
with
\begin{eqnarray}
Y_e = \left (\begin{array}{l}
V_a I_{N_{e}-1,N_{p}}(v) \\
i V_b I_{N_{e},N_{p}}(v) \\
V_c I_{N_{e}-1,N_{p}}(v) \\
i V_d I_{N_{e},N_{p}}(v) \\
\end{array} \right),
\end{eqnarray}
where $v=q^2/2eB$ with $q={(p_{{\nu}x}+p_{nx})} \hat{e}_{x} +
{(p_{{\nu}y}+p_{ny})} \hat{e}_{y}$.
The expression for  $D$ appearing in the expressions
for in $V_i$ ($i=a,b,c$ and $d$) can be
approximated by $D \approx 1/\sqrt{2}$. This leads to the following
expression for the leptonic part
\begin{eqnarray}
\mathbf{L^{\mu \nu}}=
\left (\begin{array}{cccc}
J_+(v) & 0 & 0 & J_-(v) \\
0 & J_+(v)  & i J_-(v) & 0 \\
0 & -i J_-(v) & J_+(v) & 0 \\
J_-(v) & 0 & 0 & J_+(v)
\end{array} \right)
\end{eqnarray}
with $J_{\pm}(v)= I^2_{N_{e},N_{p}}(v)
\pm I^2_{N_{e}-1,N_{p}}(v)$.
As a result the following relation is obtained for the squared norm of the interaction
matrix
\begin{eqnarray}
\sum_{R_e R_p} |{\cal M}_{R_{p},R_{e},N_{p},N_{e},s_{n},s_{p}}|^2= \frac{G^2}{2} \frac{eB}{2 \pi} (\delta_{s_{p},s_{n}}
(c^2_V + c^2_A)      \nonumber\\
(I^2_{N_{e},N_{p}}(v)+I^2_{N_{e}-1,N_{p}}(v)) \nonumber\\
+ 2
\delta_{s_p,s_n} g_a s_p
(I^2_{N_{e}-1,N_{p}}(v)-I^2_{N_{e},N_{p}}(v)) \nonumber\\
 +
2 \delta_{s_p,-s_n} g^2_a
( \delta_{s_p,1} \ I^2_{N_{e},N_{p}}(v)+ \nonumber\\
 \delta_{s_p,-1} \ I^2_{N_{e}-1,N_{p}}(v)))
 \label{urcama}
\end{eqnarray}
with $c_V=1$ and $c_A=1.26$.
\begin{acknowledgements}
      This work has been supported through
      the Stichting voor Fundamenteel  Onderzoek der Materie
      with financial support from the Nederlandse Organisatie
      voor Wetenschappelijk Onderzoek. The research of R.G.E.
      Timmermans was made possible by a fellowship of the Royal Netherlands
      Academy of Arts and Sciences.
\end{acknowledgements}

\end{document}